\documentstyle[twoside,fleqn,espcrc2,epsf]{article}

\newcommand{\nn}{\nonumber}

\newcommand{\AmS}{{\protect\the\textfont2
  A\kern-.1667em\lower.5ex\hbox{M}\kern-.125emS}}

\title{ Negative mode problem in false vacuum decay with gravity
\thanks{Talk
given at Constrained Dynamics and Quantum Gravity 99, Villasimius,
(Sardinia, Italy), September 14-18, 1999. To apper in the Proceedings.}
}

\author{George Lavrelashvili\address{Department of Theoretical Physics \\
                                     A. Razmadze Mathematical Institute \\
                                     GE-380093 Tbilisi, Georgia}%
        }

\begin{document}

\begin{abstract}
There is a single negative mode in the spectrum of small perturbations about
the tunneling solutions describing a metastable vacuum decay in flat
spacetime. This mode is needed for consistent description of decay processes.
When gravity is included the situation is more complicated. There are
different answers in different reduction schemes. An approach based on
elimination of scalar field perturbations shows no negative mode, whereas
the recent approach based on elimination of gravitational perturbations
indicates presence of a negative mode. In this contribution we analyse and
compare the present approaches to the negative mode problem in false vacuum
decay with gravity.

\vspace{0.5cm}

\rightline{gr-qc/0004025}

\vspace{0.5cm}

\end{abstract}

\maketitle

\section{INTRODUCTION}

The first order phase transitions might play an important role in the Early
Universe \cite{Linde}. They proceed via nucleation of the bubbles of true
vacuum in metastable (false) one and subsequent growth of the bubbles. The
false vacuum decay is usually discussed in frame of self-interacting scalar
field theory. This process is described by the $O(4)-$symmetric  bounce
solution \cite{Coleman,CD} of the Euclidean equations of motion. Value of the
Euclidean action at the bounce gives leading exponential factor in decay
rate. The perturbations about the bounce solution define the one loop
corrections to the bubble nucleation rate and determine quantum state of
materialized bubble \cite{CC}.  It is remarkable that in the spectrum of
small perturbations about the bounce in {\bf flat} spacetime there is exactly
one negative mode \cite{CC}.  This mode is responsible for making correction
to the ground-state energy imaginary, thus justifying decay interpretation.

When gravity is included the model contains gauge degrees of freedom and to
extract information about the spectrum one should find proper reduction to
physical variables. In the Euclidean gravity 
the presence of the gauge degrees of freedom manifests
itself in the well known conformal factor problem.
In pure gravity it is possible to cure this problem  either at the level of
prescription \cite{GHP}, or via  careful gauge fixing \cite{SH}.  For scalar
perturbations in the theory of a scalar field coupled to gravity it is
the only spatially homogeneous modes, which suffer from the conformal factor
problem (see e.g. \cite{Lav-HT}).  On the other hand the negative mode is
expected exactly in this problematic sector of scalar homogeneous
perturbations, which consists of coupled scalar field and metric
perturbations. The task is to reduce this coupled system to a system with a
single physical variable. Note that there is no problem in elimination of
unphysical variables in perturbations about the {\bf flat}
Friedmann-Robertson-Walker (FRW) type model with scalar field \cite{MFB}.
Problem arises for perturbations in the theory  of scalar field in
{\bf closed} FRW type Universe.

There are two possible ways of reduction of coupled system of
perturbations: either to eliminate metric perturbations or perturbations of
scalar field. The approach based on elimination of scalar field
perturbations shows no negative mode in the spectrum of remining variable
\cite{TS,GMST,TS2,Tanaka,Lav-HT}.
On the other hand the recent investigation \cite{KL} indicates presence of a
single negative mode in the approach based on elimination of gravitational
perturbations. This indicates on lack of full
understanding of the role of gauge-fixing procedure in description of false
vacuum decay with gravity and need for further investigation of this subject.

The aim of present contribution is to discuss and compare existing
approaches to the negative mode problem.
The rest of the article is organized as follows. In the next section we
recall main formulas relevant to false vacuum decay in flat spacetime.
In Sec. 3 we discuss inclusion of gravity, remind some important classical
solutions of Euclidean equations of motion and derive quadratic action for
scalar perturbations about $O(4)-$symmetric background solutions. We will
restrict ourselves to the most problematic sector of $O(4)-$symmetric 
perturbations.  In Sec. 4 we analyse three different reduction schemes.
Sec. 5 contains the concluding remarks.

%%%
\begin{figure}[htb]
\centerline{\epsfysize=7.5cm
\epsfbox{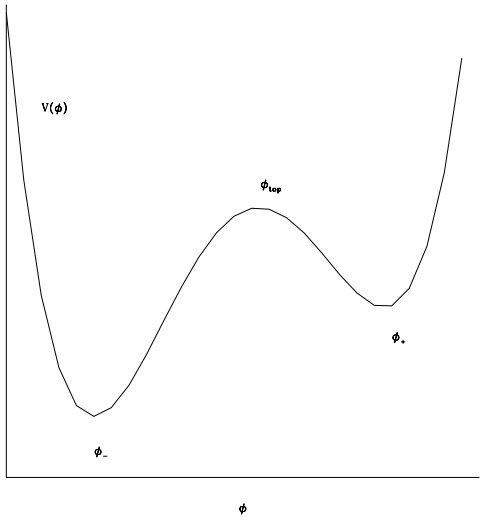}}
    \caption{Schematic view of a scalar field potential $V(\phi)$
     with a metastable vacuum $\phi=\phi_{+}$, true
    vacuum  $\phi=\phi_{-}$ and local maximum $\phi=\phi_{top}$.}
    \label{pot}
\end{figure}
%%%

\section{FALSE VACUUM DECAY IN FLAT SPACETIME}
Let's consider self-interacting scalar field $\phi$ in flat spacetime and 
assume that the scalar field potential $V(\phi)$ has shape shown in Fig.1,
with local minimum (false vacuum) at $\phi=\phi_{+}$ and absolute minimum
(true vacuum) at $\phi=\phi_{-}$.
The false vacuum decay is described by the Euclidean path integral
%%%
\begin{equation}\label{eq:efi1}
<\phi_{-}|e^{-HT}|\phi_{+}> = \int
e^{\int (i\pi_{\phi}\frac{d\phi}{d\tau}-H)d\tau}
d\phi d\pi_{\phi} \; ,
\end{equation}
%%%
where $T$ is a large positive number and  $H$ is the scalar field Hamiltonian
%%%
\begin{eqnarray}
H(\pi, \phi) = \int [\frac{\pi({\bf x})^2}{2}
+\frac{1}{2}\partial_n \phi({\bf x})\partial^n\phi({\bf x}) \nn \\
+V(\phi)] d^3 x , \; n=1,2,3.
\end{eqnarray}
%%%
Integrating Eq.(\ref{eq:efi1}) over $\pi_\phi$ one gets
%%%
\begin{equation}\label{eq:efi2}
<\phi_{-}|e^{-HT}|\phi_{+}> = \int e^{-S_E} d\phi \; ,
\end{equation}
%%%
where $S_E$ is the usual Euclidean action
%%%
\begin{equation}
S_E = \int
[\frac{1}{2}\partial_\mu \phi({x})\partial^\mu\phi({x})
+ V(\phi)] d^3 x d\tau \; ,
\end{equation}
%%%
with $\mu=0,1,2,3$.
In the limit $T\to\infty$ the l.h.s. of Eq.(\ref{eq:efi2}) contains
information about the lowest energy eigenvalue and wave function.
The r.h.s. can be calculated in the WKB approximation.
The energy $E_0$ of the lowest state gets correction due to
tunneling phenomenon
%%%
\begin{equation}\label{eq:energy}
E = E_0 - K e^{-B} \; ,
\end{equation}
%%%
with
%%%
\begin{equation}
B = S_E(\phi_{bounce})-S_E(\phi_{+}) \; ,
\end{equation}
%%%
where $\phi_{bounce}$ is the $O(4)-$ symmetric euclidean bounce solution with 
the lowest action, which interpolates between true and false vacua.
The factor $K$ is obtained by the Gaussian integration of exponent 
of quadratic action of small perturbations about the bounce 
%%%
\begin{equation}
K = \frac{B^2}{4\pi^2}
\left(\frac{det'[-\partial^2+V''(\phi)]}
{det[-\partial^2+V''(\phi_{+})]}\right)^{-1/2} \; .
\end{equation}
%%%
The operator $-\partial^2+V''(\phi)$ has zero modes and $det'$ means that
the determinant computed with the zero eigenvalues omitted.
It turns out that there is exactly one mode with the negative eigenvalue
in the spectrum of small perturbations about the bounce solution.
This makes $K$ and hence the energy shift in the Eq.(\ref{eq:energy})
purely imaginary and supports the false vacuum {\bf decay} interpretation.
Decay rate per unit volume ${\cal V}$, per unit time is given by
%%%
\begin{equation}
\frac{\Gamma}{\cal V} = |K| e^{-B} \; .
\end{equation}
%%%
Note that the functional integral Eq.(\ref{eq:efi1}) is written for
unconstrained (physical) degrees of freedom. If one has $m$
first class constraints
%%%
\begin{equation}
C_{\alpha}(\phi,\pi)\approx 0 \; , \; \alpha=1,...,m \; ,
\end{equation}
%%%
then according to general recipy \cite{SF}  one should choose some gauge
fixing conditions
%%%
\begin{equation} \chi_{\beta}(\phi,\pi) = 0 \; , \; \beta=1,...,m \; ,
\end{equation}
%%%
and modify the functional integral, Eq.(\ref{eq:efi1}), as follows
%%%
\begin{equation}
\label{eq:efi3} \int e^{\int (i\pi_{\phi}\frac{d\phi}{d\tau}-H)d\tau}
\Delta\;\delta (C_{\alpha })\;\delta (\chi^{\beta}) d\phi d\pi_{\phi} \; ,
\end{equation}
%%%
with the Faddeev-Popov determinant
%%%
\begin{equation}
\Delta\equiv \det\{C_{\alpha},\chi^{\beta}\}\neq 0 \; .
\end{equation}
%%%

\section{INCLUSION OF GRAVITY}

The Euclidean action of system composed of scalar field minimally coupled
to gravity is
%%%
\begin{equation}
S=\hskip-3pt\int d^4x \sqrt{g}\left[-\frac{1}{2\kappa} R
+ \frac{1}{2}\nabla_\mu\phi\nabla^\mu\phi + V(\phi)\right]\hskip-3pt,
\end{equation}
%%%
where $\kappa=8\pi G$ is the reduced Newton's constant.

Since we are interested in solutions having $O(4)-$symmetry,
the metric is parametrised as follows
%%%
\begin{equation} \label{eq:metric}
ds^2 = d\sigma^2 + a^2(\sigma)\gamma_{ij} dx^i  dx^j,
\; \phi=\phi(\sigma) \; ,
\end{equation}
%%%
where $\gamma_{ij}$ is the three-dimensional metric on the constant
curvature space sections.
The field equations are
%%%
\begin{eqnarray}
\ddot{a}+\frac{\kappa a}{3} ( {\dot{\phi}}^2 + V(\phi) ) = 0 ,\\
{\dot{a}}^2-{\cal K} - \frac{\kappa a^2}{3}
(\frac{{\dot{\phi}}^2}{2} - V(\phi) ) = 0 ,\\
\ddot{\varphi} + 3 \frac{\dot{a}}{a}\dot{\varphi}-
\displaystyle\frac{\delta V}{\delta \varphi} = 0,
\end{eqnarray}
%%%
where a dot denotes a derivative with respect to proper time $\sigma$
and ${\cal K}$ is the curvature parameter, which
has the values 1, 0, -1 for closed, flat and open universes respectively.
In what follows we are interested in {\bf closed} case,
but we do not specify the value of ${\cal K}$ now
in order to see how it enters in the final expressions.

\subsection{Euclidean solutions}

There are few celebrated solutions of these equations,
which play an important role in Euclidean quantum gravity.

1. The Hawking-Moss solution is a 4-sphere \cite{HM} corresponding
to scalar field sitting on the top of the potential barrier
%%%
\begin{equation} \label{eq:HM}
\phi(\sigma)=\phi_{top}, \;\;
a(\sigma) = {\cal H}_{top}^{-1} \; sin({\cal H}_{top}\sigma) \; ,
\end{equation}
%%%
with ${\cal H}_{top}=\sqrt{\kappa V(\phi_{top})/3}$. Note that the
Euclidean time $\sigma$ varies in finite interval $\sigma = (0,\sigma_{f})$.

2. The Coleman-De Luccia bounce is a deformed 4-sphere \cite{CD}.
It starts with some $\phi=\phi_{0}$ at $\sigma=0$ close to $\phi_{-}$,
stops at $\sigma=\sigma_{f}$ close to $\phi_{+}$ and obeys the
regularity conditions
%%%
\begin{equation} \label{eq:CDL}
a(0) = \dot{\phi}(0) = 0, \;\; a(\sigma_f)=\dot{\phi}(\sigma_f)=0 \; .
\end{equation}
%%%

3. Relaxing the regularity conditions
at the second zero of $a$ one gets the
Hawking-Turok singular instanton \cite{HT}.
It was introduced to describe creation of an open inflationary Universe.

Our main goal is to investigate spectrum of small perturbations about
the Euclidean solutions and, in particular,
to determine presence of a negative mode.

\subsection{Quadratic action of scalar $O(4)-$ symmetric perturbations}

Investigation of perturbations would be convenient to perform
in conformal frame.
We expand the metric and the scalar field over a $O(4)-$symmetric
background
%%%
\begin{eqnarray} \label{eq:perturb}
ds^2 & = &
a(\tau)^2\left[ (1 + 2A(\tau))d\tau^2 \right.     \nonumber \\
& + & \left. \gamma_{ij}(1-2\Psi(\tau) ) dx^i dx^j\right]  \nonumber,\\
\phi & = & \varphi(\tau) + \Phi(\tau),
\end{eqnarray}
%%%
where $\tau$ is the conformal time,
$a$ and $\varphi$ are the background field values
and $A, \Psi$ and $\Phi$ are small perturbations.

The background field equations in the conformal time are
%%%
\begin{eqnarray}
{\cal H}^2 - {\cal H}' - {\cal K} = \displaystyle\frac{\kappa}{2}
  \,\varphi{}'{}^2 ,\\
2\, {\cal H}' + {\cal H}^2 - {\cal K} =
 - \displaystyle \frac{\kappa}{2} ( \varphi'^2 + 2\, a^2 V(\varphi)), \\
\varphi{}'' + 2\, {\cal H} \varphi{}' - a^2
\displaystyle\frac{\delta V}{\delta \varphi} =0,
\end{eqnarray}
%%%
where a prime  denotes a derivative with respect to $\tau$
and ${\cal H} := a'/a$.

Expanding the total action, keeping terms of the second order in 
perturbations and using the background equations, we find
%%%
\begin{equation}
S = S^{(0)} + S^{(2)},
\end{equation}
%%%
where $S^{(0)}$ is the action for the background solution and
$S^{(2)}$ is quadratic in perturbations with the
Lagrangian for scalar $O(4)-$symmetric perturbations
%%%
\begin{eqnarray}\label{eq:lgd}
{}^{(s)}{\cal L} = \frac{1}{2 \kappa}a^2\sqrt{\gamma}
\left[ -6 \Psi'^2 + 6 {\cal K}\Psi^2 \right.          \nonumber \\
+ \kappa (\Phi'^2 + a^2
\displaystyle\frac{\delta^2 V}{\delta\varphi\delta\varphi}
\Phi^2 + 6 \varphi' \Psi' \Phi)              \nonumber \\
- (2\kappa\varphi' \Phi' - 2\kappa a^2
\displaystyle\frac{\delta V}{\delta \varphi} \Phi + 12 {\cal H} \Psi' +
12{\cal K}\Psi ) A                \nonumber \\
\left. - 2 ({\cal H}' + 2 {\cal H}^2 + {\cal K} ) A^2 \right].
\end{eqnarray}
%%%
Note that the variation with respect to $A$ gives the first order
(constraint) equation
%%%
\begin{eqnarray}\label{eq:constr}
2\kappa\varphi' \Phi'
- 2\kappa a^2 \displaystyle\frac{\delta V}{\delta \varphi} \Phi
+ 12 {\cal H} \Psi' + 12{\cal K}\Psi   \nonumber \\
+ 4 ({\cal H}' + 2 {\cal H}^2 + {\cal K} ) A = 0 \; .
\end{eqnarray}
%%%

\section{THREE APPROACHES TO THE REDUCTION}

To obtain the unconstrained system corresponding to
the degenerate Lagrangian (\ref{eq:lgd})
we will follow the conventional Dirac formulation
of generalized Hamiltonian dynamics \cite{DiracL,HenTeit}.
Calculating the canonical momenta
%%%
\begin{eqnarray}
&&\hskip-2.1em\Pi_\Psi :=i \displaystyle \frac{\delta ^{(s)}{\cal L}}{\delta \Psi'}
  = i \frac{6\,a^2\,\sqrt{\gamma}}{\kappa}\left(-\Psi' +
   \frac{\kappa}{2}\varphi'\Phi - {\cal H}\, A \right),\nonumber \\
&&\hskip-2.1em\Pi_{\Phi} := i \displaystyle \frac{\delta^{(s)}{\cal L}}{\delta \Phi'}
  = i a^2\,\sqrt{\gamma}\left(\Phi' - \varphi'\, A\right),\nonumber\\
&&\hskip-2.1em\Pi_A := i \displaystyle \frac{\delta^{(s)}{\cal L}}{\delta A'} = 0 \; ,
\end{eqnarray}
%%%
we find the primary constraint \( C_1 := \Pi_A = 0 \).

Thus the evolution is governed by the total Hamiltonian
\begin{equation}
H_T=H_C+ u_1 (\tau ) C_1 \; ,
\end{equation}
with arbitrary function \(u_1 (\tau )\) and
canonical Hamiltonian
%%%
\begin{eqnarray}\label{eq:hamcan}
&&\hskip-2em H_C = -\frac{\kappa}{12\,a^2\sqrt{\gamma}}\Pi_\Psi^2
+ \frac{1}{2\,a^2\sqrt{\gamma}}\Pi_\Phi^2
+ \frac{i \kappa \varphi'}{2} \,\Pi_\Psi\Phi             \nonumber \\
&&\hskip-1em+ a^2\sqrt{\gamma} \left[ \frac{3 {\cal K}}{\kappa} \Psi^2
+ \frac{1}{2} \left(
a^2 \displaystyle\frac{\delta^2 V}{\delta\varphi\delta\varphi}
+ \frac{3}{2}\kappa \varphi'^2 \right) \Phi^2 \right]   \nonumber \\
&&\hskip-1em+ A [i \varphi'\Pi_{\Phi} - i {\cal H} \Pi_\Psi          \nonumber \\
&&\hskip-1em+ a^2 \sqrt{\gamma} \left(
( a^2\displaystyle\frac{\delta V}{\delta \varphi}
- 3\varphi'{\cal H}) \Phi - \frac{ 6{\cal K} }{\kappa} \Psi\right) ] .
\end{eqnarray}
%%%
Conservation of primary constraint gives the secondary constraint
%%%
\begin{eqnarray}\label{eq:C2}
C_2&=& i \varphi'\Pi_{\Phi} - i {\cal H} \Pi_\Psi \nonumber \\
&+& a^2 \sqrt{\gamma} \left(
( a^2\displaystyle\frac{\delta V}{\delta \varphi}
- 3\varphi'{\cal H} ) \Phi - \frac{6{\cal K}}{\kappa} \Psi \right) .
\end{eqnarray}
%%%
The primary and secondary constraints are first class one
and there are no ternary constraints.

Under the infinitesimal shift $\tau \to \tau + \lambda$
these constraints generate the gauge transformations
%%%
\begin{eqnarray}
\delta\Psi &=&- {\cal H} \lambda \; , \qquad
\delta\Pi_{\Psi} =
\displaystyle\frac{-i 6a^2\sqrt{\gamma} {\cal K}}{\kappa}\lambda \; ,\nn \\
\delta\Phi &=&\varphi{}' \lambda \; ,\qquad
\delta\Pi_{\Phi}= i a^2\sqrt{\gamma}(\varphi{}''
-\varphi{}'{\cal H})\lambda \;, \nn \\
\delta A &=&\lambda{}'+{\cal H} \lambda \; .
\end{eqnarray}
%%%

The existence of constraints in the system as usually
means the presence of unphysical degrees of freedom and that
the whole system has no unique dynamics.
To find the physical variables with unique evolution
one should fix the gauge and solve the constraints.
There are two possible general strategies: either to eliminate
the gravitational degrees of freedom ($\Pi_{\Psi}$ and $\Psi$) or
perturbations of scalar field ($\Pi_{\Phi}$ and $\Phi$).

\subsection{The I approach}

In the first investigation on this subject the Lagrangian approach was
used \cite{LRT}. The gauge was fixed by the condition
%%%
\begin{equation} \label{eq:LRTgauge}
\Psi=0 \; .
\end{equation}
%%%
Using this gauge condition and eliminating $A$ with the help of
the constraint equation Eq.(\ref{eq:constr}) we obtain the
unconstrained quadratic action in the form
%%%
\begin{eqnarray}\label{eq:LRTaction}
S^{(2)}_{LRT}  = \int \frac{a^4 \sqrt{\gamma}}{2 Q_{LRT}} \left[
\frac{{\cal H}^2}{a^2}{\Phi'^2} -
\displaystyle\frac{\kappa\varphi'}{3} \frac{\delta V}{\delta \varphi}
\Phi'\Phi \right.             \nn \\
 +  \left. \left( \displaystyle\frac{\kappa a^2}{6}
\displaystyle\frac{\delta V}{\delta \varphi}
+ Q_{LRT}\displaystyle\frac{\delta^2 V}{\delta\varphi\delta\varphi}
\right) \Phi^2 \right] d\tau d^3 x ,
\end{eqnarray}
%%%
with
%%%
\begin{equation}
Q_{LRT}:=({\cal H}'+2 {\cal H}^2 + {\cal K} )/3
%\left({\cal K}-\displaystyle\frac{\kappa a^2}{3} V \right) 
= {\cal H}^2 - \frac{\kappa{\varphi'}^2}{6} \; .
\end{equation}
%%%

We see that the action Eq.(\ref{eq:LRTaction}) has correct overall sign
if $Q_{LRT}>0$. As it was shown in \cite{LRT} under certain choice of
parameters the bounce can develop a region with negative $Q_{LRT}$.
The main conclusion in \cite{LRT} was that the region with $Q_{LRT}<0$
leads to catastrophic particles creation and is pathological.

However the gauge fixing Eq.(\ref{eq:LRTgauge}) has problems where $a'=0$.
This can be seen in the Hamiltonian formalism,
while the Faddeev-Popov determinant is proportional to $a'$.

\subsection{The II approach}

Eliminating $\Pi_\Phi$ using the constraint Eq.(\ref{eq:C2}) and
working in the gauge invariant variables
%%%
\begin{eqnarray} \label{eq:TSvariables}
{\bf \Psi} &=&\Psi + \displaystyle\frac{\cal H}{\varphi'} \Phi, \nn \\
{\bf \Pi_{\Psi}}&=&\Pi_\Psi +
\frac{6 i {\cal K} a^2\sqrt{\gamma}}{\kappa\varphi{}'} \Phi \; ,
\end{eqnarray}
%%%
one obtains the reduced hamiltonian
%%%
\begin{eqnarray}\label{eq:hamII}
H^{\ast} = -\frac{\kappa}{12\,a^2\sqrt{\gamma}}{\bf \Pi_\Psi}^2
+ \frac{ a^2\sqrt{\gamma} 3 {\cal K}}{\kappa} {\bf \Psi^2} \nn \\
- \frac{2 a^2\sqrt{\gamma}}{{\varphi'}^2}
\left( \frac{3{\cal K}}{\kappa} {\bf \Psi}
+ \displaystyle\frac{i {\cal H}}{2 a^2\sqrt{\gamma}}
{\bf \Pi_\Psi} \right)^2 \; .
\end{eqnarray}
%%%
Performing the canonical transformation
%%%
\begin{eqnarray}\label{eq:cantransII}
{\bf \Psi}=\frac{i\kappa\varphi'}{4} \tilde{q}
-\frac{\cal H}{3{\cal K}a^2 \sqrt{\gamma} \varphi'}\tilde{p} \; , \nn \\
{\bf \Pi_{\Psi}}=\frac{3{\cal K}a^2 \sqrt{\gamma} \varphi'}{2 {\cal H}}
\tilde{q}-\frac{2 i}{\kappa\varphi'}\tilde{p} \; ,
\end{eqnarray}
%%%
and solving for the momenta $\tilde{p}$ one obtains \cite{GMST,Lav-HT}
the quadratic part of the Euclidean action
%%%
\begin{equation} \label{eaction}
{S}^{(2)} = {(1-4{\cal K})\over 2}
\int\Bigl[ ({dq\over d\tau})^2 + U q^2 \Bigr]
\sqrt{\gamma} d^3 x d\tau \;,
\end{equation}
%%%
with a potential $U$ depending on the background fields
%%%
\begin{equation}
U={\kappa\over 2}{\varphi'}^2
+{\varphi'} ({1\over{\varphi'}})''+1-4{\cal K} \;.
\end{equation}
%%%
and $q=a\tilde{q}$.

We see that quadratic action for the homogeneous harmonic
has ``wrong" overall sign.
To overcome this problem it was suggested \cite{TS} that analytic continuation
%%%
\begin{equation} \label{rotation}
q\to -i q
\end{equation}
%%%
is performed while integrating over this mode (compare \cite{GHP,RS}).

The equation for the mode functions,
which diagonalize the action (\ref{eaction}),
has form of the Schr\"odinger equation
%%%
\begin{equation} \label{setau}
-{d^2\over d\tau^2}q+U q=E q \; .
\end{equation}
%%%

Note that the singularities $\displaystyle\frac{1}{\varphi'}$,
which appear in this approach  do not allow one to investigate
straightforwardly the perturbations about the Hawking-Moss
solution (with $\varphi=\varphi_{top}=const$).

The spectrum about the Coleman-De Luccia bounce was investigated and 
no negative mode theorem was proven \cite{TS2,Tanaka}.

The Hawking-Turok instanton was investigated \cite{Lav-HT} and it was 
found that for monotonous scalar field potentials there is no negative mode
in the spectrum.
On the other hand it was found that the negative mode is present in
``exotic" cases when the Hawking-Turok instanton ``overshoots" the 
Coleman-De Luccia bounce. This means that in this case  
the Hawking-Turok instanton is unstable and has larger
action then the corresponding Coleman-De Luccia bounce.

Note that the formulation in terms of the gauge invariant variables
Eq.(\ref{eq:TSvariables}) corresponds to the gauge choice
$\chi_{\small GMST}:=\Phi=0$ (compare \cite{GMST}).
Hence the Faddeev-Popov determinant is~$\varphi'$ and this approach
is singular for certain configurations.

\subsection{The III approach}

Having in mind that
{\it i}) spherically symmetric gravity has no propagating degrees of freedom
and {\it ii}) in the limit $\kappa\to 0$ we should recover scalar field
theory in flat space, we prefer an approach based on elimination of
gravitational perturbations \cite{KL}.
At the same time from the  Eq.(\ref{eq:C2}) we see that solving constraint
we divide either to $\varphi'$ or to ${\cal H}$ or to $a^2$. Since
$\varphi'$ and $a'$ might have zero(s) by solving constraints we might
introduce extra singularities in quadratic action.  For closed universe the
scale factor $a$ is vanishing for some $\tau=\tau_{\pm}$. Since we are
interested in perturbations which vanish for $\tau\to\tau_{\pm}$, the
coordinate singularity $a(\tau_{\pm})=0$ looks most harmless.

We use the following gauge fixing conditions
%%%
\begin{equation}\label{eq:KLgauge}
\chi_1:=A=0\,, \quad \quad \chi_2 :=
\frac{\kappa}{6{\cal K} a^2\sqrt{\gamma}} \Pi_\Psi = 0 \; ,
\end{equation}
%%%
which obey the following canonical algebra:
%%%
\begin{eqnarray} \label{eq:algebra}
\{ C_i, C_j\} = \{\chi_i, \chi_j\} = 0, \quad
\{\chi_i, C_j \} = \delta_{ij}.
\end{eqnarray}
%%%
As a next step we consider all constraints in a strong sense
and define the physical Hamiltonian as
%%%
\begin{equation}
H^\ast:=H_C \big\vert_{\chi_i=0, C_i=0} \; .
\end{equation}
%%%
For the physical Hamiltonian we find
%%%
\begin{eqnarray} \label{eq:ham*}
H^\ast & = & \frac{1}{2\,a^2\sqrt{\gamma}}
\left(1-\displaystyle\frac{\kappa}{6{\cal K}}\varphi'^2 \right)
\Pi_\Phi^2                                               \nn \\
& + & \displaystyle\frac{i\kappa\varphi'}{6{\cal K}}
(a^2\displaystyle\frac{\delta V}{\delta \varphi}-3\varphi'{\cal H})
\Pi_\Phi \Phi                                            \nn \\
& + & \frac{1}{2} a^2\sqrt{\gamma} \left[
\displaystyle\frac{\kappa}{6{\cal K}}
(a^2\displaystyle\frac{\delta V}{\delta \varphi}
- 3\varphi'{\cal H})^2         \right.                   \nn \\
& + & \left .\left( a^2 \displaystyle\frac{\delta^2 V}
{\delta\varphi\delta\varphi} + \frac{3}{2}\kappa \varphi'^2 \right)
\right] \Phi^2 \; ,
\end{eqnarray}
%%%
which corresponds to the following  unconstrained quadratic action for one
physical dynamical degree of freedom
%%%
\begin{eqnarray} \label{eq:redact}
&&\hskip-2em S^{(2)}= \int
\frac{a^2 \sqrt{\gamma}}{2 Q} \left[ {\Phi'^2} -
\displaystyle\frac{\kappa\varphi'}{3{\cal K}}
(a^2\displaystyle\frac{\delta V}{\delta \varphi}
- 3 \varphi'{\cal H})\Phi'\Phi \right.             \nn \\
&& + \left( \displaystyle\frac{\kappa}{6{\cal K}}
(a^2\displaystyle\frac{\delta V}{\delta \varphi}
 - 3\varphi'{\cal H})^2  \right.                   \nn \\
&& + Q \left. \left.
\left( a^2 \displaystyle\frac{\delta^2 V}{\delta\varphi\delta\varphi}
+\frac{3}{2}\kappa \varphi'^2 \right) \right) \Phi^2 \right]
d\tau d^3 x ,
\end{eqnarray}
%%%
with
%%%
\begin{equation}
Q:= \left(1-\displaystyle\frac{\kappa}{6{\cal K}}\varphi'^2 \right) \; .
\end{equation}
%%%
In what follows we will consider the background configurations for which
the factor $Q$ is positive definite. 

Introducing a new variable $q=a/\sqrt{Q} \; \Phi$ and 
integrating by parts we obtain the unconstrained quadratic action in the form
%%%
\begin{equation}\label{eq:eucact}
S^{(2)} =\frac{1}{2} \int \left( q'^2 + W[a(\tau), \varphi(\tau)] q^2
\right) d\tau \sqrt{\gamma}d^3 x\; ,
\end{equation}
%%%
with frequency $W$ whose time dependence is determined by the background
solutions
%%%
\begin{eqnarray}\label{eq:Lpot}
&&W [a(\tau), \varphi(\tau)] = 
\frac{a^2}{Q}\displaystyle\frac{\delta^2 V}{\delta\varphi\delta\varphi}
+\frac{\kappa a^4}{2 {\cal K} Q^2}
(\displaystyle\frac{\delta V}{\delta \varphi})^2                  \nn \\
&-&\frac{2\kappa \varphi' a^2 {\cal H}} {{\cal K} Q^2} 
\displaystyle\frac{\delta V}{\delta \varphi}      
-\frac{10 {\cal H}^2}{Q}+\frac{12 {\cal H}^2}{Q^2} \nn \\
& +&\frac{8}{Q} {\cal K} -6 {\cal K} -3 {\cal K} Q  \; .  
\end{eqnarray}
%%%

Note that the obtained action allows us to consider the 
slowly varying scalar field, $\varphi' \to 0$, 
and vanishing gravity, $\kappa\to 0$, limits.

The equation for the mode functions, which diagonalize the action
Eq.(\ref{eq:eucact}), has form of the Schr\"odinger equation
%%%
\begin{equation}\label{eq:schrod}
-\frac{d^2}{d\tau^2} q + W [a(\tau), \varphi(\tau)] q = E q \; ,
\end{equation}
%%%
with the potential Eq.(\ref{eq:Lpot}).
Having this equation, one can investigate the negative mode problem for
concrete cases numerically. 
We investigated the case with the scalar field potential
%%%
\begin{equation}\label{eq:potential}
V(\varphi)=\frac{m^2}{2} (\varphi^2(\varphi-v)^2+B \varphi^4)\; .
\end{equation}
%%%
This potential\footnote{$\kappa=1$ units are used.}
for $m^2=2, B=0.12$ and $v=0.5$ has local maximum
at $\varphi_{top}=0.31250$, and local minimum (false vacuum)
at $\varphi_{false}=0.3571428$.
For this potential there exists the Coleman-De Luccia bounce solution
\cite{BL} with $\varphi_0=0.1123579$.
The quantum mechanical Schr\"odinger equation Eq.(\ref{eq:schrod})
was solved numerically for this potential and exactly one bound state
was found.
Note that the factor $Q$ was positive definite for this case.

Within this approach it is problematic to investigate the Hawking-Turok
instanton, while scalar field  $\varphi$ 
runs away and eventually the factor $Q$ becomes negative.

We found that the Schr\"odinger equation Eq.(\ref{eq:schrod}) for the 
Hawking-Moss background solution has six boundstates.
This number is in perfect agreement with the analytic formula 
for the eigenvalues \cite{TS}:
%%%
\begin{equation}\label{eq:HMspectrum}
\lambda_n=n(n+3){\cal H}_{top}^2+V''(\varphi_{top}), \; n=0,1,2...
\end{equation}
%%%
while for our example
%%%
\begin{equation}
\frac{-V''(\varphi_{top})}{{\cal H}_{top}^2}=40.96 \; .
\end{equation}
%%%
Presence of many negative modes about the Hawking-Moss solution
means that the corresponding Coleman-De Luccia bounce for the given 
potential and set of parameters gives dominant contribution to 
tunneling \cite{JS,TS}.

Note that this approach based on gauge fixing Eq.(\ref{eq:KLgauge})
can be equally well formulated  in the gauge invariant variables
%%%
\begin{eqnarray} \label{eq:KLvariables}
{\bf \Phi} &=&\Phi - \displaystyle\frac{i\kappa\varphi'}
{6a^2\sqrt{\gamma}{\cal K}}\Pi_{\Psi},  \nn \\
{\bf \Pi_{\Phi}}&=&\Pi_\Phi + \frac{\kappa}{6 {\cal K}}
(\varphi{}''-\varphi'{\cal H} )\Pi_{\Psi} \; .
\end{eqnarray}
%%%

\section{CONCLUDING REMARKS}

There are two important types of the Euclidean solutions in flat spacetime: 
instantons and bounces. 
The instanton describes quantum-mechanical mixing between the equal
energy states, whereas the bounce describes decay of a metastable state.
There are no negative modes in the spectrum of perturbations about the
instantons (only zero modes). The bounce has a single negative mode.
When gravity is included plenty of new euclidean solutions are found.
Presence or absence of a negative mode might be a good indicator for
understanding to which type each solution belongs: instanton (mixing) or
bounce (decay).

The approach based on elimination of gravitational degrees of freedom shows
no negative mode in the spectrum. It was suggested \cite{TS} that the $i$
which comes from the conformal rotation Eq.(\ref{rotation}) plays the same
role as a negative mode. This argument has a weak point.
The conformal rotation Eq.(\ref{rotation}) is the same for all euclidean
background solutions, while the imaginary shift in energy is peculiarity
of an unstable state.
Moreover, for the Giddings-Strominger wormhole a negative mode was
found on top of conformal rotation \cite{RS}.

The approach based on elimination of gravitational degrees of freedom 
indicates the presence of a negative mode.
It has smooth $\kappa\to 0$ limit and can be straightforwardly
used for the Hawking-Moss solution (with $\varphi\equiv 0$). 
At the same time there are still many open
questions. A detailed numerical study of different cases is needed.  Another
issue which needs further investigation is an understanding of the role of
configurations with the negative factor $Q$: is it real physical instability
or it is the breakdown of the reduction scheme.

\vspace{0.5cm}
I am deeply thankful to the organizers of QG99 for  splendid hospitality and
for creating charming atmosphere for discussions during the meeting.

\vspace{0.5cm}
{\it Note added:}
After this contribution was essentially completed,
further progress in investigation of negative mode problem was made.
The results are summarized in the revised version of \cite{KL}.

%
%%%
\end{document}